\documentstyle[aps,prl]{revtex}
\draft
\begin{document}
%@twocolumnfalse\endcsname
\draft\pagenumbering{roma}
\author{Liu Yu-Xi and Sun Chang-Pu}
\address{ Institute of Theoretical Physics, Academia Sinica, P.O.Box 2735,
Beijing 100080, China}
\title{Antibunching effect of the radiation field in a microcavity with 
a mirror undergoing heavily damping oscillation }
\maketitle
\thispagestyle{empty}
\vspace{15mm}
\begin{abstract}
The interaction between the radiation field in a  microcavity with 
a mirror undergoing damping oscillation is investigated. Under the heavily
damping cases, the mirror variables are adiabatically eliminated.
 The the stationary conditions of the system are discussed. 
The small fluctuation approximation around steady values is applied to
 analysis the antibunching effect of the cavity field. The antibunching 
 condition is given under two limit cases. 
\end{abstract}
\vspace{4mm}
 
\vspace{3.cm}
%\vskip2pc]
\pagenumbering{arabic}

\begin{center}{\bf 1. Introduction} \end{center} 
There has been interest in the investigation of the non-classical behavior
 of the light field, for example, the squeezed state of the electromagnetic 
 field~\cite{a,b,c,d}.
 A single mode or a multimode electromagnetic field is squeezed when the 
 fluctuation of its one quadrature component is reduced to below  the
 standard quantum limit. But the fluctuation of its conjugate partner 
 will be enlarged in terms of the uncertainty relation. The squeezed 
 electromagnetic field could be applied to the optical communication
 and ultrasensitive gravitational wave detection.
 
 Recently, it has been shown that a cavity with free oscillating mirror
 might be employed as a model for squeezing. It was firstly proposed by 
 Stenholm without taking into account the effect of the  fluctuation 
 on the oscillating mirror~\cite{e}. Since then, this model is generalized to the
 coupling of the system to the external world~\cite{f,g,h}. But up to now, another 
 non-classical behavior of the radiation field in a microcavity with
 a movable mirror , such as antibunching effect, isn't still studied.
 
 In this paper, we will discuss the  antibunching effect of the radiation 
 field in a microcavity with 
a mirror undergoing heavily damping oscillation. In section 2, we firstly
give a quantizing Hamiltonian including external dissipative effect. 
The  master equation of the  reduced density matrix for the system 
is given. We convert the master equation into a c-number equation 
(the Fokker-Planck equation) by using the two-mode positive  $P$ representation. Then the 
stochastic equation corresponding to the Fokker-Planck equation is obtained.
In section 3, we discuss the stability of the system under the good
cavity limit and linearize the  stochastic equation around stable values.
In section 4, antibunching effect is investigated and antibunching 
condition of the system is given. Finally we give the conclusion of
this paper. 

\begin{center}
{\bf 2. Hamiltonian and Master equation}  
\end{center}
We firstly  consider  the interaction between a single mode cavity field and a 
movable mirror. Based on reference~\cite{f,g,h}, we have a effective quantizing 
Hamiltonian
\begin{equation}
H_{0}=\hbar \omega_{c}a^{+} a+ \hbar \omega_{m}b^{+} b
-\hbar g a^{+} a (b^{+}+b)+i\hbar(E(t)a^{+}-E^{*}(t)a)
\end{equation}
where $E(t)$ is proportional to the amplitude of the external driving field of 
the cavity mode. $E^{*}(t)$ is complex conjugate of $E(t)$. $\omega_{c}$ is 
the cavity frequency. $a$($a^{+}$) are annihilate (create)  operators of the 
cavity field. $\omega_{m}$ is a frequency of the mirror.  $b$($b^{+}$) are 
annihilate (create) operators of the mirror.  
 $g=\frac{\omega_{c}}{L}(\frac{\hbar}{2m\omega_{m}})$. $m$ is a mass of the 
 mirror and $L$ is the  equilibrium cavity length.

If the cavity is bad, and the mirror is damped by the circumstance when it is 
moving, then the Hamiltonian (1) need correct to add the dassipative effect of the 
circumstance. So we have  
\begin{equation}
H=H_{0}+\hbar a^{+}\Gamma_{1}+\hbar a\Gamma^{+}_{1}+
\hbar b^{+}\Gamma_{2}+\hbar b\Gamma^{+}_{2}
\end{equation}
where $\Gamma_{1}(\Gamma^{+}_{1})$ is the reservoir operators of the cavity.
 We also model the  damping effect of the mirror as the result of the 
 interaction between the mirror and the many harmonic oscillator.  They 
 satisfy the Markovian correlation function (where for simplicity, 
 we only consider the zero temperature case):
\begin{eqnarray}
<\Gamma_{i}(t)\Gamma^{+}_{i}(t^{\prime})>&=&\gamma_{i}\delta(t-t^{\prime}) 
 \nonumber \\
 <\Gamma^{+}_{i}(t)\Gamma_{i}(t^{\prime})>&=&0
\end{eqnarray}  
with $i=1$, $2$.  $\gamma_{1}$ is the decay rate of the cavity field.
 $\gamma_{2}$ is the damping coefficient of the motion mirror. 
The master equation of the reduced density matrix for the system is 

\begin{eqnarray}
\frac{\partial \rho}{\partial t}&=&\frac{1}{i\hbar}[H_{0},\rho]
+\gamma_{1}(2a\rho a^{+}-\rho a^{+}a-a^{+}a\rho) \nonumber \\
&+&\gamma_{2}(2b\rho b^{+}-\rho b^{+}b-b^{+}b\rho)
\end{eqnarray}

We
 could  convert the operator eq.(4) into a c-number equation 
(the Fokker-Planck Equation) by using the two-mode positive  $P$
 representation~\cite{B} .
This representation ensures that the $P$ function exists as a well-behaved 
distribution  which is singular when the usual
Glauber-Sudarshan $P$  representation was applied.  That is:
\begin{eqnarray}
\frac{\partial P}{\partial t}&=&[\frac{\partial }{\partial \alpha_{1}}
(i\omega_{c}\alpha_{1}-ig \alpha_{1}\alpha^{+}_{2}-ig \alpha_{1}\alpha_{2}
+\gamma_{1}\alpha_{1}-E(t)) \nonumber \\
&+& \frac{\partial }{\partial \alpha^{+}_{1}}
(-i\omega_{c}\alpha^{+}_{1}+ig \alpha^{+}_{1}\alpha^{+}_{2}+
ig \alpha^{+}_{1}\alpha_{2}+\gamma_{1}\alpha^{+}_{1}-E^{*}(t)) \nonumber \\
&+&\frac{\partial }{\partial \alpha_{2}}
(i\omega_{m}\alpha_{2}-ig \alpha^{+}_{1}\alpha_{1}+
\gamma_{2}\alpha_{2}) \nonumber \\ 
&+&\frac{\partial }{\partial \alpha^{+}_{2}}
(-i\omega_{m}\alpha^{+}_{2}+ig \alpha^{+}_{1}\alpha_{1}+
\gamma_{2}\alpha^{+}_{2}) \nonumber  \\
 &+&\frac{\partial}{\partial \alpha_{1}}\frac{\partial}{\partial \alpha_{2}}
 ig \alpha_{1}
 -\frac{\partial}{\partial \alpha^{+}_{1}}\frac{\partial}{\partial \alpha^{+}_{2}}
 ig \alpha^{+}_{1}] P
\end{eqnarray}

where $\alpha_{1}$ and $\alpha^{+}_{1}$, and $\alpha_{2}$ and $\alpha^{+}_{2}$
are no longer complex conjugate of each other, instead they are independent
complex variables.  In terms of the Ito ruler, the Fokker-Planck  eq.(5) is 
equivalent to the following set of the stochastic equations.  
\begin{mathletters}
\begin{eqnarray}
\frac{\partial \alpha_{1}}{\partial t}
&=&
-i\omega_{c}\alpha_{1}+ig \alpha_{1}\alpha^{+}_{2}+ig \alpha_{1}\alpha_{2}
-\gamma_{1}\alpha_{1}+E(t)+\Gamma_{\alpha_{1}} \\ 
\frac{\partial \alpha^{+}_{1}}{\partial t}
&=&
i\omega_{c}\alpha^{+}_{1}-ig \alpha^{+}_{1}\alpha^{+}_{2}-
ig \alpha^{+}_{1}\alpha_{2}-\gamma_{1}\alpha^{+}_{1}+E^{*}(t)
+\Gamma_{\alpha^{+}_{1}} \\
\frac{\partial \alpha_{2}}{\partial t}&=&
-i\omega_{m}\alpha_{2}+ig \alpha^{+}_{1}\alpha_{1}
-\gamma_{2}\alpha_{2}+\Gamma_{\alpha_{2}} \\
\frac{\partial \alpha^{+}_{2}}{\partial t}&=&
i\omega_{m}\alpha^{+}_{2}-ig \alpha^{+}_{1}\alpha_{1}
-\gamma_{2}\alpha^{+}_{2}+\Gamma_{\alpha^{+}_{2}}
\end{eqnarray}
\end{mathletters}
where $\Gamma_{\alpha_{i}}$ and  $\Gamma_{\alpha^{+}_{i}}$ are Gaussian random
variables with zero mean. Their correlation functions satisfy: 
\begin{mathletters}
\begin{eqnarray}
<\Gamma_{\alpha_{1}}(t)\Gamma_{\alpha_{2}}(t^{\prime})>&=&
<\Gamma_{\alpha_{2}}(t)\Gamma_{\alpha_{1}}(t^{\prime})>=
ig\alpha_{1}\delta(t-t^{\prime})\\
<\Gamma_{\alpha^{+}_{1}}(t)\Gamma_{\alpha^{+}_{2}}(t^{\prime})>&=&
<\Gamma_{\alpha^{+}_{2}}(t)\Gamma_{\alpha^{+}_{1}}(t^{\prime})>=
-ig\alpha^{+}_{1}\delta(t-t^{\prime}) \\
<\Gamma_{\alpha_{l}}(t)\Gamma_{\alpha^{+}_{m}}(t^{\prime})>&=&
<\Gamma_{\alpha^{+}_{m}}(t)\Gamma_{\alpha_{l}}(t^{\prime})>=0
\end{eqnarray}
\end{mathletters}
with $l=1$ or $2$ and $m=1$ or $2$. The above correlation functions could
be obtained from the matrix elements of the diffusion matrix of eq.(5) by 
using  Ito ruler.

\begin{center}
{\bf  3.  Adiabatic elimination of the mirror variable and stability analysis}
\end{center} 
In the good-cavity limit, that is, the damping coefficient of the mirror 
is much larger than the decay rate of the cavity field 
($\gamma_{2} \gg \gamma_{1}$). This means that the decay time of the cavity 
field is much longer than the decay time of the mirror. So  the mirror variables
could be adiabatically eliminated from eqs.(6a-6b) . Firstly we find a 
steady-state of the mirror variables . In the case of the steady state, 
$\dot{\alpha}_{2}=0$ and $\dot{\alpha}^{+}_{2}=0$. So we have from eqs.(6c-6d) 
\begin{mathletters}
\begin{eqnarray}
-i\omega_{m}\alpha_{2}+ig \alpha^{+}_{1}\alpha_{1}
-\gamma_{2}\alpha_{2}+\Gamma_{\alpha_{2}}&=&0 \\
i\omega_{m}\alpha^{+}_{2}-ig \alpha^{+}_{1}\alpha_{1}
-\gamma_{2}\alpha^{+}_{2}+\Gamma_{\alpha^{+}_{2}}&=&0
\end{eqnarray}
\end{mathletters}
From eqs.(8a-8b), we obtain:
\begin{mathletters}
\begin{eqnarray}
\alpha_{2}&=&\frac{\Gamma_{\alpha_{2}}+ig\alpha_{1}\alpha^{+}_{1}}
{\gamma_{2}+i\omega_{m}} \\
\alpha^{+}_{2}&=&\frac{\Gamma_{\alpha^{+}_{2}}-ig\alpha_{1}\alpha^{+}_{1}}
{\gamma_{2}-i\omega_{m}} 
\end{eqnarray}
\end{mathletters}
 We substitute eqs.(9a-9b) into eqs.(6a-6b) and eliminate the mirror variables. 
 Then we have:
\begin{mathletters}
\begin{eqnarray}
\frac{\partial \alpha_{1}}{\partial t}  
&=&-i\omega_{c}\alpha_{1}+iG(\omega_{m})\alpha^{+}_{1}\alpha^{2}_{1}
-\gamma_{1}\alpha_{1}+\Gamma \\
\frac{\partial \alpha^{+}_{1}}{\partial t}  
&=&i\omega_{c}\alpha^{+}_{1}-iG(\omega_{m})\alpha^{+2}_{1}\alpha_{1}
-\gamma_{1}\alpha^{+}_{1}+\Gamma^{+} 
\end{eqnarray}
\end{mathletters}
with $G(\omega_{m})=\frac{2g^{2}\omega_{m}}{\gamma_{2}+\omega^{2}_{m}}$ and
\begin{mathletters}
\begin{eqnarray}
\Gamma&=&\frac{ig\alpha_{1}}{\gamma_{2}+i\omega_{m}}\Gamma_{\alpha_{2}}+
\frac{ig\alpha_{1}}{\gamma_{2}-i\omega_{m}}\Gamma_{\alpha^{+}_{2}}
+\Gamma_{\alpha_{1}}\\
\Gamma^{+}&=&-\frac{ig\alpha^{+}_{1}}{\gamma_{2}+i\omega_{m}}\Gamma_{\alpha_{2}}-
\frac{ig\alpha^{+}_{1}}{\gamma_{2}-i\omega_{m}}\Gamma_{\alpha^{+}_{2}}
+\Gamma_{\alpha^{+}_{1}}
\end{eqnarray}
\end{mathletters}
Using eqs.(7a-7c) and eqs.(11a-11b), we calculate the correlation function
\begin{mathletters}
\begin{eqnarray}
<\Gamma(t)\Gamma^{+}(t^{\prime})>&=&<\Gamma^{+}(t)\Gamma^{+}(t)>=
\frac{2\gamma_{2}g^{2}n}{\gamma^{2}_{2}+\omega^{2}_{m}}\delta(t-t^{\prime}) \\
<\Gamma^{+}(t)\Gamma^{+}(t^{\prime})>&=&-
\frac{2g^{2}\alpha^{+2}_{1}}{\gamma_{2}-i\omega_{m}}\delta(t-t^{\prime}) \\
<\Gamma(t)\Gamma(t^{\prime})>&=&-
\frac{2g^{2}\alpha^{2}_{1}}{\gamma_{2}+i\omega_{m}}\delta(t-t^{\prime})
\end{eqnarray}
\end{mathletters}
The eqs.(10a-10b) are difficult to be solved. In general, we are interested in
the properties of the steady state. So we denote the steady values of 
$\alpha_{1}$ and $\alpha^{+}_{1}$ by $\alpha_{0}$ and $\alpha^{+}_{0}$ 
respectively. We assume that the system has a small approximation around
the steady values, namely
\begin{equation}
\left \{ \begin{array}{l}
\alpha_{1}(t)=\alpha_{0}+\delta\alpha_{1}(t)\\
\alpha^{+}_{1}(t)=\alpha^{+}_{0}+\delta\alpha^{+}_{1}(t)
\end{array} \right.
\end{equation}
So non-linear eqs.(10a-10b) are simplified into the following linear 
equations around the steady values. 
\begin{eqnarray}
\frac{\partial}{\partial t} \left ( \begin{array}{c} \delta\alpha_{1} \\ 
\delta\alpha^{+}_{1} \end{array} \right )&=&
\left ( \begin{array}{cc} 
-i\omega_{c}-\gamma_{1}+i2G(\omega_{m})\alpha^{+}_{0}\alpha_{0} & 
iG(\omega_{m})\alpha^{2}_{0} \\
-iG(\omega_{m})\alpha^{+2}_{0}& 
i\omega_{c}-\gamma_{1}+i2G(\omega_{m})\alpha^{+}_{0}\alpha_{0}
\end{array} \right ) 
\left ( \begin{array}{c}\delta\alpha_{1} \\ 
\delta\alpha^{+}_{1} \end{array} \right )
 \nonumber \\
&+& 
\left ( \begin{array}{cc}
-\frac{2g^{2}\alpha^{2}_{0}}{\gamma_{2}+i\omega_{m}}
&  \frac{2\gamma_{2}g^{2}n}{\gamma^{2}_{2}+\omega^{2}_{m}}\\ 
\frac{2\gamma_{2}g^{2}n}{\gamma^{2}_{2}+\omega^{2}_{m}}
&-\frac{2g^{2}\alpha^{+2}_{0}}{\gamma_{2}-i\omega_{m}}
\end{array} \right )^{\frac{1}{2}}
\left ( \begin{array}{c} \eta_{1}(t) \\ \eta_{2}(t)  \end{array} \right ) 
\end{eqnarray}
where $n=\alpha^{+}_{0}\alpha_{0}$ and $\eta_{i}(t)$ statisfy delta correlation
 function:
\begin{mathletters}
\begin{eqnarray}
<\eta_{i}(t)>&=&0 \\
<\eta_{i}(t)\eta_{j}(t^{\prime})>&=&\delta_{ij}\delta(t-t^{\prime})
\end{eqnarray}
\end{mathletters}
We abbreviate eq.(8) as following:
\begin{equation}
\frac{\partial}{\partial t}\delta \vec{\alpha}(t)=-A\vec{\alpha}(t)+
D^{\frac{1}{2}}\vec{\eta}(t)
\end{equation}

But one of the important feature of eq.(14) is whether the steady solutions
are stable, that is,  when $\alpha_{1}$ and $\alpha^{+}_{1}$ somewhat deviate 
from $\alpha_{0}$ and $\alpha^{+}_{0}$, whether they will still return to steady  
values. So we neglect the fluctuation forces of eq.(16) and have:
\begin{equation}
\frac{\partial}{\partial t}\delta \vec{\alpha}(t)=-A\vec{\alpha}(t)
\end{equation}
In order to investigate the stationary of the system,  we seek the 
solutions of the form $e^{-\lambda t}$ of eq.(17). The eigenvalues $
\lambda$ are determined by the equation
\begin{equation}
|A-\lambda I|=0
\end{equation}
where $I$ is a identity matrix. The solutions of the eq.(18) are
\begin{equation}
\lambda=\gamma_{1}\pm \sqrt{(\omega_{c}-3G(\omega_{m})n)
(\omega_{c}-G(\omega_{m})n)}
\end{equation}
This equation indicates if 
$ G(\omega_{m})n \leq \omega_{c} \leq 3G(\omega_{m})n$, the real part
of the eigen-solution of the eq.(18) is positive. The system is 
stable. When $\omega_{c}\leq n G(\omega_{m})$ or 
$\omega_{c}\geq 3n G(\omega_{m})$, only if 
$\gamma_{1} \geq \sqrt{(\omega_{c}-3G(\omega_{m})n)
(\omega_{c}-G(\omega_{m})n)}$ then the system is also stable.
So only we chose the proper parameter, the system always may
reach to stability. The small fluctuation approximation is appreciate.

\begin{center}
{\bf 4. Second-order correlation function}
\end{center}
The intensity correlation is another quantity of the experimental interest  
besides the first order optical coherence. It's truly photon correlation 
measurement. Theoretically, people have defined a second-order correlation
function to investigate the joint photocount probability of detecting the
arrival of a photon at one time and other ptoton at another time. For 
zero time delay, The second order correlation function is~\cite{cc}  
\begin{equation}
g^{(2)}(0)=\frac{<a^{+}a^{+}aa>}{<a^{+}a>^{2}}
\end{equation}
Under the positive $P$ representation  we keep the second order terms of 
$\frac{\delta\alpha}{\alpha}$, then eq(20) becomes into:
\begin{equation}
g^{(2)}(0)==\frac{<\alpha^{+2}\alpha^{2}>}
{<\alpha^{+}\alpha>^{2}}=1+\frac{<\delta\alpha^{2}>}{\alpha^{2}}+
\frac{<\delta\alpha^{+2}>}{\alpha^{+2}}+
4\frac{<\delta\alpha^{+}\delta\alpha>}{\alpha^{+}\alpha}
\end{equation}

In order to calculate the correlation functions  $<(\delta\alpha)^{2}>$ et.al,
 we need calculate matrix:
\begin{equation}
C=\left ( \begin{array}{cc}
<\delta\alpha \delta\alpha> &<\delta\alpha \delta\alpha^{+}> \\
<\delta\alpha^{+}\delta\alpha> & <\delta\alpha^{+} \delta\alpha^{+}>
\end{array} \right )
\end{equation}

The  matrix $C$ could be obtained by calculating of $\delta\alpha_{1}$ and 
$\delta\alpha^{+}_{1}$ from the linearized eq.(14). It is also could 
be given by~\cite{C}:
\begin{equation}
C=\frac{D.Det(A)+(A-ITr(A))D(A-ITr(A))^{T}}{2Tr(A)Det(A)}.
\end{equation}

after tedious calculation, we have:
\begin{mathletters}
\begin{eqnarray}
C_{11}&=&C^{*}_{22}=\frac{\alpha^{2}_{1}[2M\gamma^{2}_{1}+2NG(B+i\gamma_{1})-i2\gamma_{1}BM
-G^{2}n^{2}(M+M^{*})]}{4\gamma_{1}(\gamma^{2}_{1}+B^{2}-G^{2}n^{2})} \\
C_{12}&=&C_{21}=\frac{2N(\gamma^{2}_{1}+B^{2})+i\gamma_{1}Gn^{2}(M^{*}-M)
-GBn^{2}(M+M^{*})}{4\gamma_{1}(\gamma^{2}_{1}+B^{2}-G^{2}n^{2})}
\end{eqnarray}
\end{mathletters}
with
\begin{equation}
\left.\begin{array} {ccc}
M=-\frac{2g^{2}}{\gamma_{2}+i\omega_{m}} &N=\frac{2\gamma_{2}g^{2}n}
{\gamma^{2}_{2}+\omega^{2}_{m}} & B=\omega_{c}-2Gn
\end{array} \right.
\end{equation}

So the second order correlation function is:
\begin{equation}
g(0)=1+\frac{G[(\omega_{c}-2nG)(2n\gamma_{2}\omega_{c}-n^{2}\gamma_{2}G
-\gamma_{1}\omega_{m})+n(\gamma^{2}_{1}\gamma_{2}+n^{2}G^{2}\gamma_{2}
-2\gamma_{1}\omega_{m}Gn)]}
{n\gamma_{1}\omega_{m}[\gamma^{2}_{1}-(\omega_{c}+5nG)(\omega_{c}-nG)]}
\end{equation}
From this equation, we see that the antibunching of the cavity field 
 appear when the second term of the eq.(26) is a negative number. This condition 
could be satisfied after we chose some proper parameters. Now we 
consider two limit cases. In the case of the strong cavity field, we only 
keep the terms including $n^{3}$. The second term of eq.(26) is positive, no 
antibunching appears. But under case of the weak cavity field, namely 
when $n \rightarrow 0$, the cavity field presents antibunching behavior.

\begin{center}
{\bf 5. Conclusion}
\end{center}
The interaction between the radiation field in a  microcavity with 
a movable mirror undergoing heavily damping oscillation is investigated. 
Under the heavily damping cases, the mirror variables are adiabatically 
eliminated. The stationary condition of the system is given.
The small fluctuation approximation around steady values is applied to 
analysis the antibunching behavior of the cavity field. In the case of 
the strong cavity field, no antibunching appears. But 
under the case of the weak cavity field, cavity field presents
the antibunching effect.  

\end{document}